\documentclass[prb,aps,twocolumn,showpacs,floats]{revtex4}
\usepackage[dvips]{graphicx}
\usepackage{epsfig,bm,graphics}
\usepackage{epsfig}
\usepackage[francais]{babel}
\usepackage[utf8]{inputenc}
\usepackage{amsmath,amssymb}

\begin{document}
\title{Monte Carlo Study of Magnetic Resistivity in Semiconducting MnTe}
\author{Y. Magnin and H. T. Diep\footnote{ Corresponding author, E-mail:diep@u-cergy.fr }}
\address{ Laboratoire de Physique Th\'eorique et Mod\'elisation,
Universit\'e de Cergy-Pontoise, CNRS, UMR 8089\\
2, Avenue Adolphe Chauvin, 95302 Cergy-Pontoise Cedex, France.}
\begin{abstract}
We investigate in this paper properties of the spin resistivity in the magnetic semiconducting MnTe of NiAs structure. MnTe is a crossroad semiconductor with a large band gap.  It is an antiferromagnet with the N\'eel temperature around 310K. Due to this high N\'eel temperature, there are many applications using its magnetic properties. The method we use here is the Monte Carlo simulation in which we take into account the interaction between itinerant spins and lattice Mn spins. Our results show a very good agreement with experiments on the shape of the spin resistivity near the N\'eel temperature.
\end{abstract}
\maketitle

\section{Introduction}

Spin resistivity in materials has been a subject of intensive studies both experimentally and theoretically for more than five decades.
 Experiments have been performed to determine the spin resistivity $\rho$ in  many magnetic materials from metals to semiconductors.   The rapid development of the field is due mainly to many applications in particular in spintronics.  One interesting aspect of magnetic materials is the existence of a magnetic phase transition from a magnetically ordered phase to the paramagnetic (disordered) state. Depending on the material, $\rho$ can show a sharp peak at the magnetic transition temperature $T_C$,\cite{Matsukura} or just only a change of its slope, or an inflexion point.  The latter case gives rise to a peak of the differential resistivity $d\rho/dT$.\cite{Stishov,Shwerer}  Very recent experiments such as those performed on ferromagnetic SrRuO$_3$ thin films\cite{Xia}, Ru-doped induced ferromagnetic La$_{0.4}$Ca$_{0.6}$MnO$_3$\cite{Lu},  antiferromagnetic $\epsilon$-(Mn$_{1-x}$Fe$_x$)$_{3.25}$Ge\cite{Du}, semiconducting Pr$_{0.7}$Ca$_{0.3}$MnO$_3$ thin films\cite{Zhang}, superconducting BaFe$_2$As$_2$ single crystals\cite{Wang-Chen}, and La$_{1-x}$Sr$_x$MnO$_3$\cite{Santos}   compounds show different forms of anomaly of the magnetic resistivity at the magnetic phase transition temperature.

 The properties of the total resistivity stem  from different kinds of diffusion processes: the scattering of the itinerant spins by phonons, by lattice magnons, by impurities and defects etc.
Each contribution has in general a different temperature dependence.  Let us summarize the most important contributions at low temperature ($T$) in the following expression
\begin{equation}
\rho(T)=\rho_0+AT^2+BT^5+C\ln{\frac{\mu}{T}}
\end{equation}
where $A$, $B$ and $C$ are constants. The first term is $T$-independent,  the second term proportional to $T^2$ represents the scattering of itinerant spins at low $T$ by lattice spin-waves. Note that the resistivity caused by a Fermi liquid is also proportional to $T^2$.   The $T^5$ term corresponds to low-$T$ resistivity in metals.  This is due to the scattering of itinerant electrons by phonons.  Note that at high $T$, metals show a linear-$T$ dependence.  The $\ln$ term is the resistivity due to the quantum Kondo effect at very low $T$.
For the magnetic contribution to the total resistivity, the $T^2$ term has been obtained from the magnon scattering by Kasuya\cite{Kasuya}.  However, at high $T$ in particular in the region of the phase transition, much less has been known.  The general idea that the magnetic resistivity  is a function of
 the spin-spin correlation was introduced by de Gennes and Friedel.\cite{DeGennes}
According to this idea, the magnetic resistivity should behave as the magnetic susceptibility, thus it should diverge at $T_C$.
Fisher and Langer\cite{Fisher}, and
Kataoka\cite{Kataoka} have suggested that the range of spin-spin correlation changes  the shape of $\rho$ near the phase transition.  The resistivity due to magnetic impurities has been calculated by Zarand et al.\cite{Zarand} as a function of the Anderson's localization length.  This parameter expresses in fact a kind the correlation sphere induced around each impurity.  Their result that the resistivity peak depends on this parameter is in agreement with the spin-spin correlation idea.
 In our previous works\cite{Akabli,Akabli2,Akabli3} we have studied the spin current in ferromagnetic thin films by Monte Carlo (MC) simulations. The behavior of the spin resistivity
 as a function of  $T$ has been shown to be in agreement with main experimental features and theoretical investigations mentioned above.  We have introduced in these works the picture of scattering of itinerant spins by magnetic defect clusters which are known to be formed in the transition temperature region.  The size of each defect cluster expresses a kind of correlation between spins.

 In antiferromagnets much less is known because there have been very few theoretical
investigations which have been carried out.
 Haas\cite{Haas} has shown that while in ferromagnets the resistivity $\rho$   shows a sharp peak at the magnetic transition of the lattice spins, in antiferromagnets there is no such a peak.
 The alternate change of sign of the spin-spin correlation with distance may have something to do with the absence of a sharp peak.

In this paper, we are interested in  the antiferromagnetic MnTe,  a well-studied semiconductor with numerous applications due to its high N\'eel temperature.   The pure MnTe has either the zinc-blend structure\cite{Hennion} or the hexagonal NiAs one\cite{Hennion2}.   We confine ourselves in the case of hexagonal structure. For this case, the N\'eel temperature is $T_N=310$ K\cite{Hennion2}.
Hexagonal MnTe is a crossroad semiconductor with a big gap (1.27 eV) and a room-temperature carrier concentration of
$n=4.3\times 10^{17}$cm$^{-3}$.\cite{Mobasser,Allen}  Without doping, MnTe is non degenerate.  In doped cases\cite{Szwacki,Komatsubara,Wei,Adachi}, band tails created by doped impurities can cover more or less the gap.   But these systems, which are disordered by doping, are not a purpose of our present study.  So, in the following we study only the pure MnTe.  The behavior of the spin resistivity $\rho$ in MnTe
 as a function of $T$ has been experimentally  shown\cite{Chandra,Li,Russe,Efrem,He}.  In our previous paper,\cite{Akabli4} we have theoretically studied with the Boltzmann's equation using numerical data for cluster distribution obtained by MC simulations.  We could only compare our result with old experimental data available at that time below the transition temperature.\cite{Chandra} After the publication of our paper, new experimental results become available for the whole range of temperature for pure MnTe performed by He et al.\cite{He} This motivates the present work.

In this paper, we use the same model as in our previous work\cite{Akabli4}  for  MnTe.  But unlike that previous paper which used an approximate Boltzmann's equation, we shall use here direct MC simulations  without recourse to approximations. In addition, we take into account the lattice magnetic relaxation time in the calculation of the spin resistivity. The tuning of the relaxation time allows an excellent agreement with experimental resistivity of MnTe as will be seen below.

In section \ref{model}, we show our model and describe our MC method.  Results are shown and compared with experimental resistivity of MnTe in section \ref{result}.
Concluding remarks are given in section \ref{conclu}.

\begin{figure}[h!]
 \centering
 \includegraphics[width=60mm,angle=0]{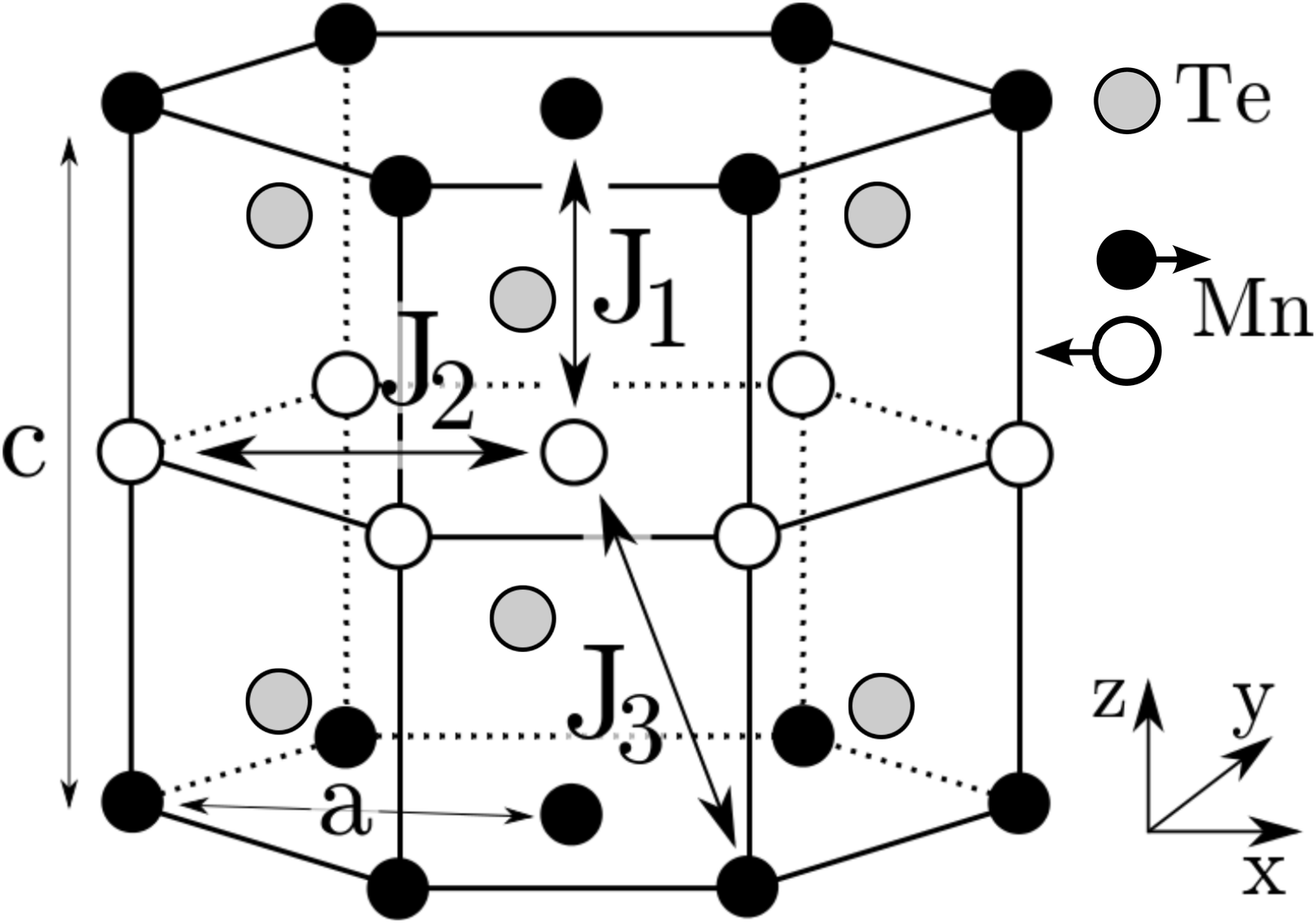}
 \caption{Structure of MnTe of  NiAs type  is shown. Antiparallel spins are shown by black and white circles.  Nearest-neighbor (NN) interaction is marked by $J_1$, next NN interaction by $J_2$, and third NN one by $J_3$.} \label{NiAs}
\end{figure}

\section{Model}\label{model}
The hexagonal NiAs-type structure of MnTe is shown in Fig. 1.\cite{Hennion2} It is composed of ferromagnetic $xy$ hexagonal planes antiferromagnetically stacked in the $c$ direction.  The nearest-neighbor (NN) distance in the $c$ direction is  $c/2\simeq 3.36$ $\AA$ shorter than the in-plane NN distance which is $a=4.158 \AA$.
Neutron scattering experiments show that the main exchange interactions between Mn spins in MnTe
are i) interaction between NN along the $c$ axis with the value  $J_1/k_B=-21.5\pm 0.3$ K,  ii)  ferromagnetic exchange
$J_2/k_B \approx 0.67\pm 0.05$ between in-plane  neighboring Mn (they are next NN by distance), iii)  third NN antiferromagnetic interaction $J_3/k_B\simeq -2.87\pm 0.04$ K.

The interactions $J_1$, $J_2$ and $J_3$ are indicated in Fig.~\ref{NiAs}.  In addition, the spins are lying in the $xy$ planes perpendicular to the $c$ direction with a small in-plane easy-axis anisotropy $D$.\cite{Hennion2}
We note that the values of the exchange integrals given above have been deduced from experimental data by fitting with a formula obtained from a free spin-wave theory\cite{Hennion2}. Other fittings with mean-field theories give slightly different values.\cite{Mobasser}   Therefore, care should be taken while using values deduced from experimental data, keeping in mind that they depend on models used and approximations involved in the fitting.

The lattice Hamiltonian is given by

\begin{eqnarray}
{\cal H} &=& -J_1\sum_{(i,j)} \mathbf{S}_i.\mathbf{S}_j -J_2\sum_{(i,m)} \mathbf{S}_i.\mathbf{S}_m
-J_3\sum_{(i,k)} \mathbf{S}_i.\mathbf{S}_k  \nonumber \\
&& -D\sum_i (S_i^x)^2 \label{HL}
\end{eqnarray}
where $\mathbf{S}_i$ is the Heisenberg spin at the lattice site $i$, $\sum_{(i,j)}$ is made
over the NN spin pairs $\mathbf{S}_i$ and $\mathbf{S}_j$ with  interaction $J_1$, while $\sum_{(i,m)}$ and
$\sum_{(i,k)}$ are made over the NNN and third NN neighbor pairs with interactions $J_2$ and $J_3$, respectively.  $D>0$ is an anisotropy constant which favors the  in-plane $x$ easy-axis spin configuration.
  The Mn spin is experimentally known to be of
the Heisenberg model with  magnitude $S=5/2$.\cite{Hennion2}

The interaction between  an itinerant spin  and surrounding Mn spins  in  semiconducting MnTe  is written as
\begin{equation}\label{interact}
{\cal H}_i=-\sum_nJ(\vec r-\vec R_n)\mathbf  s\cdot \mathbf S_n
\end{equation}
where $J(\vec r-\vec R_n)>0$ is a ferromagnetic exchange interaction between the itinerant spin $\mathbf s$ at $\vec r$ and the Mn spin $\mathbf S_n$ at the lattice site $\vec R_n$ .
The sum on lattice spins $\mathbf S_n$
is limited at some cut-off distance as will be discussed later.
We suppose that $J(\vec r-\vec R_n)$ is weak enough to be
considered as a perturbation to the lattice Hamiltonian:

\begin{equation}\label{interact1}
J(\vec r-\vec R_n)=I_0\exp [-\alpha(\vec r-\vec R_n)]
\end{equation}
where $I_0$ and $\alpha$ are constants. We choose $\alpha=1$ for convenience.  The choice of $I_0$ should be made so that the interaction ${\cal H}_i$ yields an energy much smaller than the lattice energy due to ${\cal H}$ (see discussion on the choice of variables in Refs. \cite{Magnin,Magnin2}).

Since in MnTe the carrier concentration is
 $n=4.3\times 10^{17}$cm$^{-3}$, low with respect to the concentration of its surrounding lattice spins
$\simeq 10^{22}$cm$^{-3}$,  we do not take into account the interaction between itinerant spins.  The contribution of that interaction to the energy of an itinerant spin is negligible.

We perform MC simulations on a thin film of  dimension $L\times L \times L$ where $L$ is the number of MnTe cells in $x$, $y$ and $z$ directions. Note that each cell contains two Mn atoms and two Te atoms (see Fig. \ref{NiAs}).  Periodic boundary conditions are applied in  all directions.   The itinerant electrons move in the system under an electric field $\vec{\epsilon}$ applied along the $x$ direction
\begin{equation}
\mathcal{H}_E  =  -e\vec{\epsilon}.\vec{\ell}
\end{equation}
where $-e$ is the charge of electron and $\vec \ell$ the displacement vector of the electron.   Each electron spin $\mathbf s$ of magnitude $1/2$ interacts with neighboring lattice Mn spins within a sphere of radius $D_1$ according to Eq. (\ref{interact}).  We take in the following the Ising model for the electron spin. In doing so, we neglect the quantum effects which are of course important at very low temperature but not in the transition region at room temperature where we focus our attention.

The MC technique we use for the transport is a multi-step averaging which has shown to reduce efficiently statistical fluctuations\cite{Magnin,Magnin2}. The reader is referred to our early works for a detailed description of our method as well as the effects of changing physical parameters such as $D_1$ in antiferromagnets.  In a word, this technique consists in changing the lattice spin configuration as often as the temperature-dependent relaxation time allows.  The relaxation time of the lattice spin system  is expressed as\cite{Hohenberg}
\begin{equation}\label{tau}
\tau_L=\frac{A}{|1-T/T_N|^{z\nu}}
\end{equation}
where $A$ is a constant, $\nu$  the correlation critical exponent, and $z$ the dynamic exponent.  From this expression, we see that as $T$ tends to $T_N$, $\tau_L$ diverges.  This phenomenon is known as the critical slowing-down.  For the Heisenberg model, $z\nu=1.38$ ($\nu=0.704$ and $z=1.97$).\cite{Peczak}   We have previously shown  that $\tau_L$ strongly  affects the shape of  $\rho$.\cite{Magnin3}

\section{Results}\label{result}

We have calculated the spin resistivity of  the hexagonal MnTe using the exchange integrals taken from Ref. \cite{Hennion2}. These values are given above.
As said before, the values of the exchange interactions deduced from experimental  data depend on the model Hamiltonian, in particular the spin model, as well as the approximations.  Furthermore, in semiconductors, the carrier concentration is a function of  $T$.
In our model, there is however no interaction between itinerant spins. Therefore, the number of itinerant spins used in the simulation is important only for statistical average: the larger the number of itinerant spins the better the statistical average.  The current obtained is proportional to the number of itinerant spins but there are no extra physical effects.

\begin{figure}[h!]
 \centering
 \includegraphics[width=80mm,angle=0]{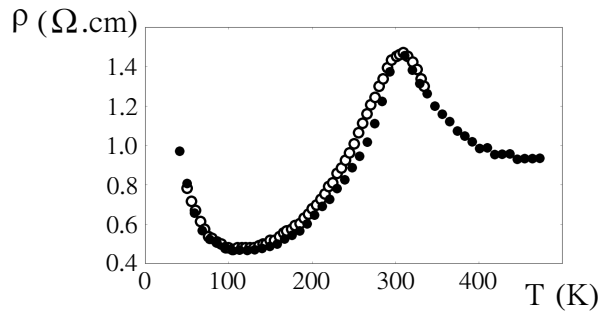}
 \caption{Spin resistivity $\rho$ versus temperature $T$. Black circles are from Monte Carlo simulation, white circles are experimental data taken from He et al.\cite{He}. The parameters used in the simulation are $J1=-21.5$K, $J_2=2.55$ K, $J_3=-9$ K, $
I_0=2$ K, $D=0.12$ K, $D_1=a=4.148\AA$, $E = 2*10^5$ V/m, $L=30$.} \label{RMnTe}
\end{figure}

The result of the spin resistivity is shown in
Fig.~\ref{RMnTe}.  Using the value of $\rho$, we obtain the relaxation time  of itinerant spin equal to $\tau \simeq 0.1$ ps, and the mean free path equal to $\bar l \simeq  20 \AA$, at the critical temperature.

Several remarks are in order:

i) With $J_3$ slightly larger in magnitude than the value deduced from experiments, we find $T_N=310$ K

ii) $\rho$ shows a pronounced peak in excellent agreement with experiments

iii) Note that the shape of the peak depends on $A$.  The value we used to obtain that agreement is $A=1$

iv) In the temperature regions below $T<140$ K and above $T_N$ the MC result is in excellent agreement  with experiment, unlike in our previous work\cite{Akabli4} using the Boltzmann's equation

v) In the region $140$ K $<T<T_N$ the MC result of $\rho$  is slightly smaller than experimental data. In the search for an explanation, we see that the magnetization obtained by the MC simulation, though in good agreement with experiments at low $T$ and yielding the precise value of $T_N$,  is slightly smaller than the experimental one in the intermediate temperature region. This is shown in Fig. 3.  The magnetization deficit may be due to the fact that the magnetic anisotropy was too small $D=0.12$ K  taken from Ref. \cite{Hennion2}, it is not strong enough to enhance the magnetization.

\begin{figure}[h!]
 \centering
 \includegraphics[width=80mm,angle=0]{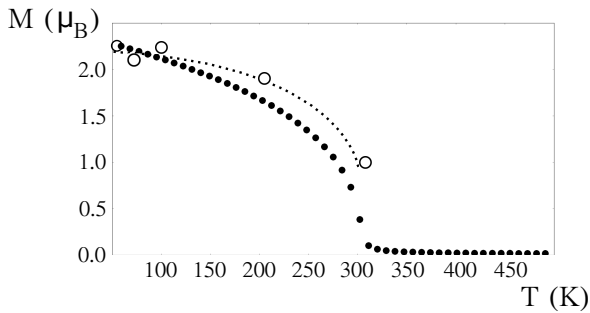}
 \caption{Magnetization $M$ (in unit of $\mu_B$) versus temperature $T$ (in unit of K). Black circles are results of Monte Carlo simulation, white circles are experimental data taken from Efrem D'Sa et al.\cite{Efrem}. The parameters are the same as those given in the caption of Fig. 2.} \label{MT}
\end{figure}

 \section{Conclusion}\label{conclu}

 We have shown in this paper the MC results of the spin resistivity
 $\rho$ as a function of temperature in MnTe.  We have taken into account the main interaction which governs
 the resistivity behavior, namely the interaction
 between itinerant spins and the lattice Mn spins.   Our result
 is in agreement with experiments: it reproduces the correct N\'eel temperature as well as the shape of the peak at the phase transition.  Note that the theory by Haas\cite{Haas}  predicts the absence of a peak in $\rho$ in the
 temperature region of the phase transition for antiferromagnetic MnTe.   We finally emphasize that our excellent agreement was possible  because most importantly we have correctly taken into account the temperature dependence of the lattice spin relaxation time in the simulation.\cite{Magnin3}

{}

\end{document}